\documentclass[aps,prl,twocolumn,showpacs,superscriptaddress]{revtex4-1}

\usepackage{amsmath,amsfonts,amssymb}
\usepackage{graphicx}
\usepackage{dcolumn}

\begin{document}


\title{Path Integral Calculations of the Hydrogen Hugoniot Using Augmented
Nodes}


\author{Saad A. Khairallah}
\email{khairallah1@llnl.gov}
\affiliation{Lawrence Livermore National Laboratory, Livermore, CA 94550}
\author{J. Shumway}
\email{john.shumway@asu.edu}
\homepage{http://shumway.physics.asu.edu}
\affiliation{Department of Physics, Arizona State University, Tempe AZ 85287}
\author{Erik W. Draeger}
\affiliation{Lawrence Livermore National Laboratory, Livermore, CA 94550}


\date{\today}

\begin{abstract}
We calculate the hydrogen Hugoniot using ab initio path integral Monte
Carlo. We introduce an efficient finite-temperature fixed-node
approximation for handling fermions, which includes an optimized
mixture of free particle states and atomic orbitals. The calculated
Hugoniot confirms previous fixed-node path integral calculations at
temperatures around $T=30\,000$~K and above, while approaching
smoothly the low temperature gas gun results. The ability to optimize
the free energy within the path integral opens many new possibilities
for developing nodal density matrices for path integral simulations of
other chemical systems.
\end{abstract}


\maketitle

Predicting the properties of materials at extreme conditions of
high pressures and high temperatures is a fundamental challenge in computational
physics, particularly when quantum effects are significant.  One such
system is the hydrogen Hugoniot, measured in laser
shock-wave~\cite{DaSilva:1997,Collins:1998} and gas
gun~\cite{Nellis:1983} experiments, which has become one of the
benchmarks for state-of-the-art theory.

Path Integral Monte Carlo (PIMC) allows direct calculation of
thermodynamic properties of many-body quantum systems at finite
temperature~\cite{Ceperley:1995,Pierleoni:1994,Magro:1996,Militzer:2000,Militzer:2001a,Militzer:2009,Hu:2010}.  For fermions, an additional
``fixed-node'' constraint is imposed to prevent an exponential
slowdown in convergence with increasing system size and decreasing
temperature~\cite{Ceperley:1992,Ceperley:1996}.  This requirement to
supply an \emph{a priori} form for the nodal surface of the many-body
wavefunction has limited the application of PIMC almost exclusively to
hydrogen and helium systems at high temperatures.  Not only does the
use of an approximate nodal function introduce uncontrolled error,
traditional algorithms for enforcing the nodal constraint have been
shown to become increasingly inefficient at lower temperatures,
ultimately failing to satisfy the Monte Carlo ergodicity requirement.
For hydrogen, nodal errors begin to manifest
at lower temperatures, below 100\,000~K~\cite{Militzer:2000},
and ergodicity becomes difficult to achieve below 20\,000~K~\cite{Militzer:2000}.

In this Letter, we present a simplifying approximation to fixed-node
PIMC that resolves several of these issues.  This new implementation
is fully symmetric in imaginary time, thus achieving better
computational efficiency in addition to avoiding the ergodicity issues
of the traditional approach.  More importantly, the method allows the
relative free energies of different nodal choices to be sampled
directly within PIMC simulations, \emph{thus providing a systematic
way to construct an optimal nodal function}.  This is an important
independent verification of other ab initio proposals and brings the
benefits of PIMC to a much broader physical domain. While the ability
to test and improve nodes using total energies has been common in
ground-state Quantum Monte Carlo (QMC), extending practical tests of
nodes to compare free energies is an important advancement in the
study of finite temperature nodes.  This could have far-reaching
consequences, as atomic and molecular orbitals can be introduced
without bias, potentially paving the way to extend PIMC to heavier
elements and lower temperatures.

\begin{figure}[tl]
\centerline{\includegraphics[width=0.92\linewidth]{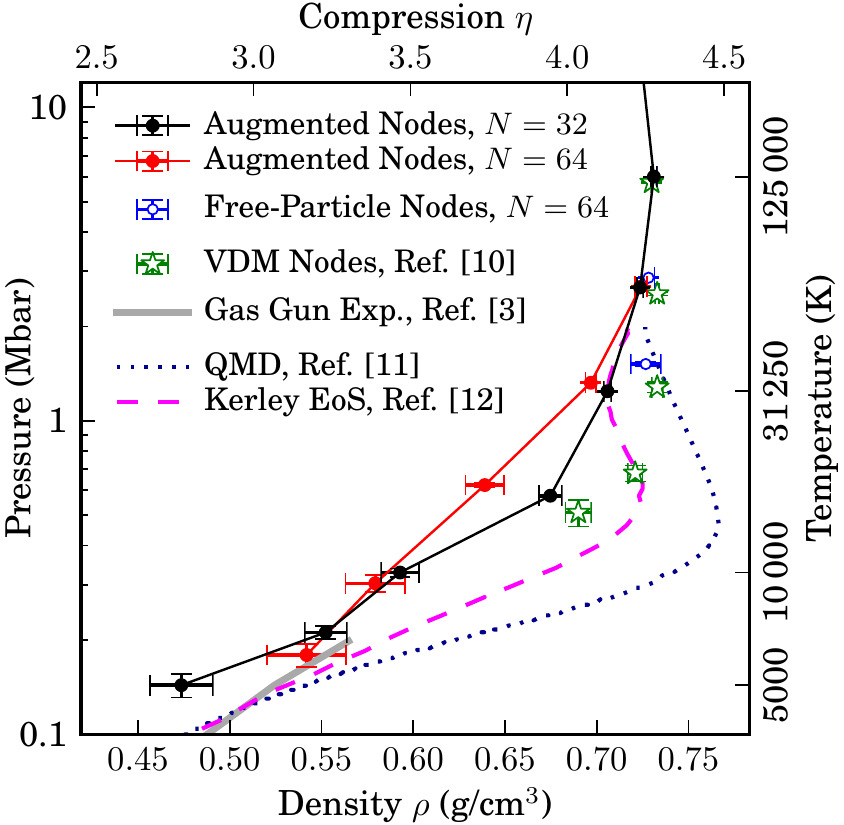}}
\caption{(color online) The Deuterium Hugoniot curve calculated within PIMC from
$T=125\,000$~K down to $T=5\,000$~K. While free particle nodes are
sufficient at high temperatures (above 125\,000~K, see
Table~\ref{table:energy} ), augmented nodes are necessary once bound
electronic states form. Our Hugoniot curve approaches the
correct free particle limit of fourfold compression in the high
temperature and pressure limit (not shown).
\label{fig:hugoniot}}
\end{figure}

We demonstrate the new formalism with calculations of the hydrogen
Hugoniot (Fig.~\ref{fig:hugoniot}), using augmented nodes
incorporating 1s atomic orbitals.  Our calculations of the hydrogen
Hugoniot span temperatures from $T=1\,000\,000$~K all the way down to
$5\,000$~K, overlapping for the first time with ground state theories
like Density Functional Theory (DFT)~\cite{Desjarlais:2003,Bonev:2004},
as well as lower-temperature gas
gun experiments~\cite{Nellis:1983}. There is a general consensus over the high and low
pressure limits of the Hugoniot curve. A controversy arises at the
intermediate pressure, around 1~Mbar, where laser-driven shock
measurements \cite{DaSilva:1997} observe high compression
$\eta\sim5.5-6$ while magnetically driven flyers \cite{Knudson:2001}
and convergent explosives \cite{Belov:2002} find a stiffer curve
closer to $\eta \sim 4$. We agree with the latter with a maximum
compression measurement $\eta \sim 4.2$.
We see no evidence of a plasma phase transition that has been
predicted by some other simulations \cite{Filinov:2001,Bezkrovniy:2004}.
Our results are consistent with recent QMC and DFT calculations that show that a 
metal-insulator phase transition occurs only stakes place at lower temperatures,
below 2\,000~K, at pressures above 1.2~Mbar \cite{Morales:2010,Lorenzen:2010}.

\begin{figure}[tl]
\begin{center}
\includegraphics[width=3.375in]{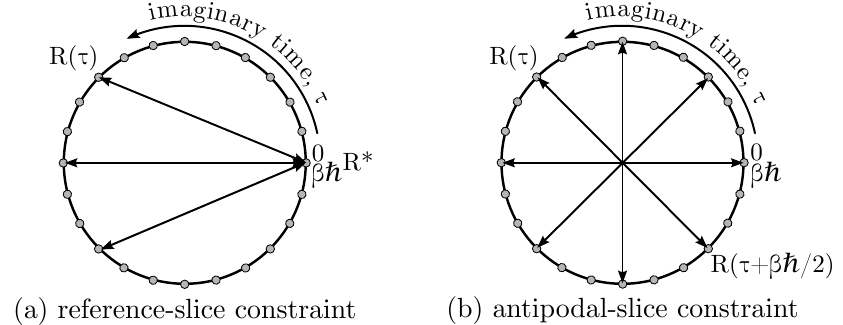}
\caption{Schematic illustration of the (a) reference-slice fixed-node, 
Eq.~(\ref{eq:refSlice}),~\cite{Ceperley:1992} and (b) 
antipodal-slice fixed-node, Eq.~(\ref{eq:antipodal}), 
constraints for path integrals.}
\label{fig:diagram}
\end{center}
\end{figure}

PIMC uses Feynman's path integral formulation of statistical mechanics
to describe a quantum  system in thermal equilibrium with a
path integral over imaginary time:
\begin{equation}\label{eq:pathIntegral}
\langle R | e^{-\beta H} |R'\rangle
= \int_{\scriptsize
\begin{array}{c}R(\beta\hbar)=R \\R(0)=R'\end{array}
} \mathcal{R}(\tau) e^{-\frac{1}{\hbar}S_E[R(\tau)]},
\end{equation}
where $R$ represents all the particle coordinates, and the path $R(\tau)$
extends for an imaginary time $\beta\hbar = \hbar/ k_B T$.
In PIMC simulations, the path is discretized into $N_T$ slices,
where $N_T$ is the Trotter number, which divides imaginary time
into short imaginary-time intervals, $\Delta\tau=\beta\hbar/N_T$.
To handle fermions, the path integral, Eq.~(\ref{eq:pathIntegral}),
is anti-symmetrized over all permutations of identical particles,
which leads to an exponential decrease in algorithmic efficiency 
at low temperatures and large system sized, known as the
fermion sign problem.

One strategy to manage the fermion sign problem is the fixed-node approximation.
The traditional implementation of the fixed-node approximation restricts 
the path integral to paths that satisfy the constraint~\cite{Ceperley:1992},
\begin{equation}\label{eq:refSlice}
\rho_T(R(\tau),R^*;|\tau|) >0 \quad (\forall \tau \in  [-\beta\hbar/2 ,\beta\hbar/2]),
\end{equation}
where the reference slice coordinates are taken at $\tau=0$ so that
$R^*=R(0)$, as shown in Fig.~\ref{fig:diagram}(a).  This rigorously
enforces the nodal constraint, but adds imaginary-time dependence to
the evaluation of the nodal action~\cite{Ceperley:1992}.
The nodal model $\rho_T$ must be defined for imaginary times ranging
from $\Delta\tau$ to $\beta\hbar/2$, corresponding to temperatures
ranging from $2T$ to $N_TT$.

In the antipodal slice approach, we make a simplifying approximation,
\begin{equation}\label{eq:antipodal}
\rho_T(R(\tau),R(\tau+\beta\hbar/2) \ne 0 \quad (\forall \tau \in  [-\beta\hbar/2 ,0]),
\end{equation}
illustrated in Fig.~\ref{fig:diagram}(b).  For each slice $R(\tau)$,
we evaluate the constraint with the slice $R(\tau+\beta\hbar/2)$ that
is antipodal to the slice, rather than the reference slice $R^*$.  While we have
tested this constraint on many problems, we have not determined its
range of validity;  the results presented here are a 
pragmatic test of the antipodal approximation. By
lifting the imaginary-time dependence, the antipodal slice
approximation allows one to avoid ergodicity errors caused by the
reference slice getting ``pinned'' at lower temperatures where the
number of imaginary-time slices $N_T$ becomes large and long permutation
cycles become increasingly common.  
The nodal model is greatly simplified, as $\rho_T$ now only needs to
be evaluated at one temperature $2T$ for the imaginary time $\tau=\beta\hbar/2$.
Parallel scalability is similarly increased, as
antipodal collections of slices can simultaneously make multiple 
multilevel moves~\cite{Ceperley:1995} on
local processors without requiring global communication.

The most significant advancement allowed by the antipodal
approximation, however, is the ability to directly compare the free
energies of different nodal models.  We hypothesize that the
best nodal choice will the lowest free energy, in analogy
to the best ground state wavefunction nodes having the lowest energy.
For the nodal density matrix, we
take a product of Slater determinants,
\begin{equation}
\rho_T(R,R') = \det | \rho(\mathbf{r}_{i\uparrow},\mathbf{r}_{j\uparrow}) |\;
\det | \rho(\mathbf{r}_{i\downarrow},\mathbf{r}_{j\downarrow}) |,
\end{equation}
where $\rho(\mathbf{r},\mathbf{r})$ are single-particle density matrices
at a temperature $T/2$.
In this work we augment the single particle density matrices by
including a hydrogen 1s orbital in the thermal mixture,
\begin{equation}
\begin{aligned}
\label{eq:augNodes}
\rho(\mathbf{r},\mathbf{r}';\beta,w) = 
&\frac{\exp\left(-\frac{m|\mathbf{r}_i-\mathbf{r}'_j|^2}{2\beta\hbar^2}\right)}
{(2\pi\beta\hbar^2/m)^{3/2}}\\
+&w\sum_k^{N_{\text{ion}}}\psi_{\text{1s}}(\mathbf{r}_i-\mathbf{r}'_j)
\psi^*_{\text{1s}}(\mathbf{r}'_j-\mathbf{r}'_k)
\end{aligned}
\end{equation}
where the weight $w$ is a free parameter and 
$\psi_{\text{1s}}(\mathbf{r})$ is the normalized hydrogen 1s orbital.
Here the first term is the free particle thermal density matrix, and
the sum is a mixture of density matrices for hydrogenic orbitals
about each of the ions, each with a weight $w$.

\begin{figure}[t]
\begin{center}
\includegraphics{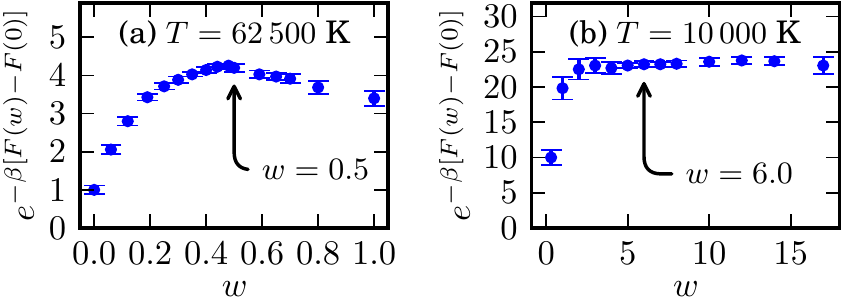}
\caption{(color online) The relative partition function 
$Z\propto\exp(-\beta F[\rho_T(R,R';T,w)])$ for the system with $r_s=2.0$,
for different nodal choices. The fixed-node free energy $F[\rho_T]$ is a functional
of the nodal density matrix, $\rho_T$. The parameter
$w$ is the weight of the 1s orbital in the augmented nodal mixture,
Eq.~(\ref{eq:augNodes}). At higher temperatures, there is a clear maximum
in the free energy, while the free energy saturates for large values of $w$ at low
temperatures. Arrows indicate optimal values of $w$, and
we report all choices of $w$ in Table~\ref{table:energy}.}
\label{fig:freeEnergy}
\end{center}
\end{figure}

To find the optimal value of $w$, we histogram free
energy differences for nodes with different values of $w$ using the acceptance ratio
method \cite{Bennet:1976},
as shown in Fig.~\ref{fig:freeEnergy}.
At $T=62\,500$~K a value of $w$ around 4.7 increases the partition function by
about a factor of four. Note that the partition function decreases for larger 
values of $w$, indicating an optimal mixture of atomic 1s and free particle
density matrices occurs near $w=5$.
At $T=10\,000$ a value of $w$ around 3 or larger increases the partition function by more
than a factor of 20. The partition function is relatively insensitive to $w$ for all
$w>3$, indicating that the 1s orbitals dominate the nodal surface.

Such a systematic approach for optimizing a given nodal model from
first principles has not been possible before within the reference
slice approach, due to the fact that the nodes are time-dependent and
require a complete set of terms over the full range of inter-slice
temperatures from $MT$ down to $2T$.
We note that this is analogous to optimizing ground nodes directly 
diffusion Monte Carlo energies, which is the most rigorous ground state
fixed-node strategy~\cite{Umrigar:2007,Reboredo:2009}.

\begin{figure}[t]
\begin{center}
\includegraphics{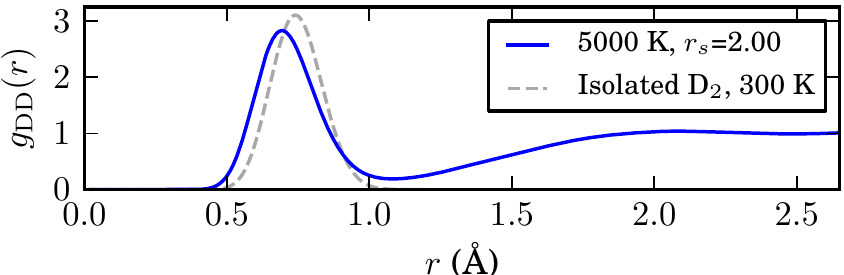}
\caption{(color online) The pair correlation function $g(r)$ for 32 deuterium atoms at 
$T=5000$~K, computed
using PIMC with augmented nodes.  The contribution
to $g(r)$ from an isolated molecule at $300~K$ is shown for comparison, illustrating 
the compression of the molecules.}
\label{fig:PCF}
\end{center}
\end{figure}


To allow direct comparison with traditional reference slice PIMC,
we performed calculations of the hydrogen Hugoniot using the same
densities as Ref.~\onlinecite{Militzer:2000}.  We ran periodic simulation
with $N=32$ and $N=64$ deuterium atoms and for both 
densities with $r_s=2.00$ and $r_s=1.86$.  Following
Ref.~\onlinecite{Militzer:2000}, we identify the Hugoniot curve by linear
interpolation. The two densities accurately bracket the
Hugoniot curve at  higher pressures, but was found to introduce
an extrapolation error at the lowest pressures, making the statistical error
bars shown in Fig.~\ref{fig:hugoniot} a lower bound only.  Following
the guidance in Ref.~\onlinecite{Militzer:2000} on the importance of small
time steps, we used $\Delta\tau/\hbar=0.02$~Ha$^{-1}$, a
time step at least twice as small as Ref.~\onlinecite{Militzer:2000}.
Our results are summarized in  Fig.~\ref{fig:hugoniot} and Table~\ref{table:energy}.

\begin{table*}[t]
\caption{PIMC calculation of deuterium Hugoniot and compressibilities with respect to an 
initial cryogenic state of $0.171$~g/cc at $T=19.6$~K. A small time step of $\tau^{-1}
=1.536\times 10^7$~K ($\Delta\tau/\hbar=0.02$~Ha$^{-1}$, twice smaller than 
Ref.~\cite{Militzer:2000}) is used.}
\begin{ruledtabular}
\begin{tabular}{rD..5D..5D..2D..5D..5D..2D..5D..5D..7}
&\multicolumn{1}{c}{$p$ (Mbar)}
&\multicolumn{1}{c}{$E$ (eV)} 
&\multicolumn{1}{c}{$w$}
&\multicolumn{1}{c}{$p$ (Mbar)}
&\multicolumn{1}{c}{$E$ (eV)}
&\multicolumn{1}{c}{$w$}
&\multicolumn{1}{c}{$p$ (Mbar)}
&\multicolumn{1}{c}{$\rho$ (g/cm$^3$)}
&\multicolumn{1}{c}{$\eta$} \\
T(K) 
&\multicolumn{1}{c}{$r_s=2$} 
&\multicolumn{1}{c}{$r_s=2$}
&\multicolumn{1}{c}{$r_s=2$}
&\multicolumn{1}{c}{$r_s=1.86$}
&\multicolumn{1}{c}{$r_s=1.86$}
&\multicolumn{1}{c}{$r_s=1.86$} 
&\multicolumn{1}{c}{Hugoniot}
&\multicolumn{1}{c}{Hugoniot}
& \multicolumn{1}{c}{Hugoniot} \\
\hline
1\,000\,000 
& 54.14(3) & 245.2(1) & 0.
& 67.35(5) & 244.6(2) & 0. 
& 56.51(3) & 0.7028(3) &4.110(2) \\
500\,000 
& 26.19(2) & 113.1(1) & 0.
& 32.46(2) & 111.9(1) & 0.
& 27.74(1) & 0.7138(4) & 4.174(2)\\
250\,000 
& 12.24(1) & 45.8(1) & 0.
& 15.18(2) & 44.8(1) & 0.
& 13.16(1) & 0.7249(6) & 4.239(4) \\
125\,000 
& 5.50(2) & 12.4(1) & 0.
& 6.94(2) & 11.9(1) & 0.
& 6.01(1) & 0.7310(2) &4.275(1) \\
62\,500 
& 2.430(19) & -3.44(8) & 0.5
& 3.232(27) & -3.29(10) & 0.5
& 2.677(15) & 0.724(4) & 4.24(2)\\
31\,250 
& 1.269(8) & -9.76(3) & 0.8 
& 1.650(17) & -10.00(6) & 0.75
& 1.322(7) & 0.697(3) & 4.08(2)\\
15\,625 
& 0.575(5) & -13.26(3) & 3.0
& 0.713(9) & -13.33(3) & 0.6 
& 0.577(5) & 0.675(6) & 3.95(4) \\
10\,000 
& 0.394(5) & -14.55(2) & 16.0
& 0.517(7) & -14.48(2) & 6.0
& 0.306(10) & 0.556(15) & 3.25(9) \\
7\,813 
& 0.336(5) & -15.04(2) & 8.5
& 0.504(7) & -15.05(3) & 8.0
& 0.211(10) & 0.552(12) & 3.23(7) \\
5\,000 
&0.219(4) & -15.72(2) & 11.0
&0.281(7) & -16.03(2) & 9.0
&0.143(12) & 0.475(18) & 2.77(11)\\
\end{tabular}
\end{ruledtabular}
\label{table:energy}
\end{table*}

 In the high temperature and pressure limit, our Hugoniot approaches
the correct free particle limit of fourfold compression. We see
excellent agreement with past PIMC results at high temperatures and
pressures ~\cite{Militzer:2000}, where permutations are infrequent and
free particle nodes accurately describe a dense strongly interacting
plasma.  An augmented node description starts becoming necessary below
$62\,500$ K, resulting in lower free energy than free particle nodes
alone.  Using augmented nodes, we see the Hugoniot curving towards
lower compression in the 1~Mbar regime, before eventually coming
closer to gas-gun and DFT low pressure
results~\cite{Nellis:1983,Caillabet:2011}.  We note that the antipodal
slice approximation also speeds up simulations considerably, as we
find that parallel simulations run more than four times faster in
direct comparisons to reference-slice fixed-node simulations.

We note that these results differ somewhat from the lowest pressure
results of Ref.~\cite{Militzer:2000}, which used a variational
approach to optimize the parameters in a density matrix assumed to be
a Slater determinant of Gaussians.  The variational nodes are shown to
give lower energies than the free particle density matrix 
in Ref.~\onlinecite {Militzer:2000}, however, no
formal proof is given that variational nodes used there result in the lowest free
energy.  Further investigation is needed to resolve this discrepancy.
We note that using 1s orbitals in our augmented node results is also a
crude model that can be enhanced by adding higher orbitals.

Our results do not seem to suffer from finite size effects at higher
temperatures. This is not the case at low temperatures. We comment on
the exceptional discrepancy at T$=15\,625$~K. This is where DFT-MD
shows some difficulty probing the simultaneous molecular bonding and
dissociations that take place in this range. DFT which can be
notorious in describing bond lengths and band gap closure could be
exaggerating a high compression trend. Long distance correlations and
delocalized electron paths are observed in our PIMC simulations. This
may suggest the need for larger PIMC simulation box size and more
atoms to account for these correlations.

We note that in the direct PIMC approach by~\cite{Bezkrovniy:2004}, the PIMC
simulations did not seem to converge to an equilibrium state
around $10\,000$~K, and a plasma transition into partially ionized
deuterium was assumed to take place in that region. A notable
configurational expression of this is the nucleii form metallic
clusters (droplets) with highly delocalized electrons over 
them~\cite{Bezkrovniy:2004}. We do not observe this behavior in our simulation.

Our data are within error bars of the magnetically driven 
flyer~\cite{Knudson:2001} and convergent explosive~\cite{Belov:2002}
experiments. If the recent correction on the equation of state of the quartz
standard~\cite{Knudson:2009} is taken into account, the laser data  
shifts to lower compressions indicating that Hydrogen is not
as compressible as the older data suggested.

In summary, we present a new time-independent formalism for fixed-node
fermionic PIMC which is significantly more efficient and free of the
ergodicity issues that plague the traditional reference slice
approach.  A clear strategy for creating nodal models with lower free
energy emerges from this approach, which we anticipate will open the
door to systems previously inaccessible to PIMC, including elements
with $Z>2$.  We present updated results on the hydrogen Hugoniot which
show a clear trend toward the lower compression and closer agreement
with Kerley's equation of state \cite{Kerley:2003}.

\begin{acknowledgments}
We are thankful for the fruitful discussions with David Ceperley,
Kevin Schmidt, Michael Bonitz, Ken Esler, Brian Clark, Jonathan
Dubois, Heather Whitley and Bernie Alder. This work was performed
under the auspices of the U.S. Department of Energy by Lawrence
Livermore National Laboratory under Contract DE-AC52- 07NA27344.  We
also acknowledge financial support from NSF DMR 02-39819.
\end{acknowledgments}

\bibliography{HydrogenPRL}

\end{document}